# CodEv: An Automated Grading Framework Leveraging Large Language Models for Consistent and Constructive Feedback

En-Qi Tseng
*Department of Information Management*
*National Sun Yat-Sen University*
Kaohsiung, Taiwan
nseventseng@gmail.com

Pei-Cing Huang
*Department of Information Management*
*National Sun Yat-Sen University*
Kaohsiung, Taiwan
pcpeicing@gmail.com

Chan Hsu
*Department of Information Management*
*National Sun Yat-Sen University*
Kaohsiung, Taiwan
chanshsu@gmail.com

Peng-Yi Wu
*Department of Information Management*
*National Sun Yat-Sen University*
Kaohsiung, Taiwan
pengyiwu0963@gmail.com

Chan-Tung Ku
*Department of Information Management*
*National Sun Yat-Sen University*
Kaohsiung, Taiwan
kuchantung@gmail.com

Yihuang Kang
*Department of Information Management*
*National Sun Yat-Sen University*
Kaohsiung, Taiwan
ykang@mis.nsysu.edu.tw

*Abstract*—Grading programming assignments is crucial for guiding students to improve their programming skills and coding styles. This study presents an automated grading framework, CodEv, which leverages Large Language Models (LLMs) to provide consistent and constructive feedback. We incorporate Chain of Thought (CoT) prompting techniques to enhance the reasoning capabilities of LLMs and ensure that the grading is aligned with human evaluation. Our framework also integrates LLM ensembles to improve the accuracy and consistency of scores, along with agreement tests to deliver reliable feedback and code review comments. The results demonstrate that the framework can yield grading results comparable to human evaluators, by using smaller LLMs. Evaluation and consistency tests of the LLMs further validate our approach, confirming the reliability of the generated scores and feedback.

I. INTRODUCTION

Introductory programming courses are the cornerstone of computer education and a necessary prerequisite for further study in advanced courses. These courses introduce students to basic programming concepts and practices; however, the abstract nature of computer concepts and the complexity of coding logic often hinder learning in the initial stages. Students require targeted guidance to correct errors, write concise code, and demonstrate how the code operates. With the proliferation of programming courses, the number of students and the grading burden on teachers are increasing, making it challenging for educators to evaluate code more efficiently and provide effective feedback.

While grading has traditionally been a manual process, recent advances in Machine Learning (ML) and Artificial Intelligence (AI) have introduced automated grading methods to simplify the grading tasks. Automated grading used to rely heavily on executing test cases to assess student submissions [1]. In this regard, ML models, such as natural language processing (NLP) techniques like recurrent neural networks (RNN), convolutional neural networks (CNN), and long short-term memory networks (LSTM) [2], have enhanced the grading process. However, these methods have an obvious limitation– they solely rely on the model's output to provide the final scores. More is needed; to further evaluate students' programming ability, we must develop methods and techniques that are able to better assess the correctness, readability, and overall structures of students' code.

With the rise of online learning and massive open online courses (MOOCs), educators face a significant challenge: providing comprehensive evaluations of numerous student code submissions within limited time constraints, covering aspects such as correctness, readability, and structural integrity. Under such immense pressure, many teachers often offer only simple scores or extremely brief comments. However, relying solely on these cursory suggestions is insufficient to foster students' ability to write high-quality code genuinely. Effective learning requires more profound and specific guidance, enabling students to understand what constitutes good programming and continually improve through practice.

The emergence of LLMs addresses the limitations of previous methods that relied solely on the program output. With their exceptional natural language understanding capabilities, LLMs enable the grading process to consider the code's readability and structures, making it possible for more complex tasks, such as generating feedback and syntax/logical error detection. For automated grading tasks, most existing research primarily centers on the performance of large-parameter models, such as GPT-3.5 and GPT-4 [3], [4], [5], [6]. However, using proprietary models (e.g., GPT-4o, Gemini [7]) increases the cost for educators, whereas applying open-weight large-parameter models to grading requires more computational resources [8], making it challenging to employ LLMs for automated grading in practical applications. Furthermore, LLMs frequently struggle to follow instructions and provide reliable, unbiased outputs [9]. Therefore, it is crucial to carefully evaluate the results generated by LLMs when using them as a basis for grading.

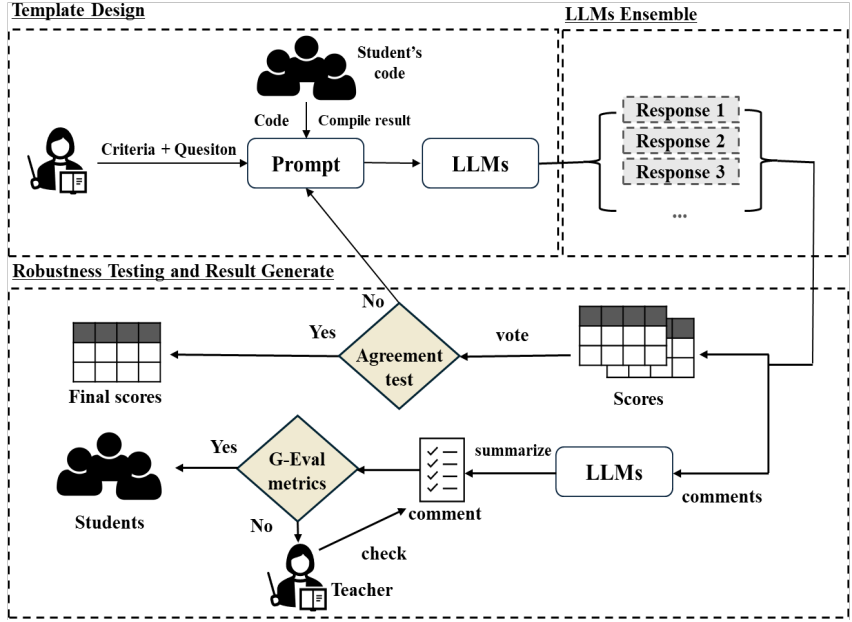

Fig. 1. CodEv framework

To overcome these limitations, this study introduces the CodEv (for "Code Evaluation"), an LLM-based code evaluation workflow/framework, which incorporates multiple dimensions of code evaluation (e.g., readability and functionality) into the grading criteria. CodEv generates scores alongside corresponding comments, allowing for a more comprehensive code assessment. We applied Chain of Thought (CoT) [10] and LLM ensemble methods to improve the performance and stability of LLMs in automated grading. The proposed CodEv also narrows the performance gap between small-parameter and large-parameter models and enables using small-parameter models in automated grading tasks.

To ensure the consistency and reliability of LLM-generated responses, we conducted both intra-model and inter-model consistency tests, ensuring stable and accurate evaluation results. Specifically, we evaluate the quality of the code feedback with the G-Eval framework [11], a metrics-based approach for LLMs, to investigate the generated feedback's results further.

In summary, the main contributions of this paper are as follows:

- We propose an LLM-based code evaluation workflow/framework that assesses code readability and structure while generating constructive comments.
- We demonstrate that the proposed LLM-based framework performs well in terms of output (both grades/scores and review comments/feedback) consistency and reliability on grading, by utilizing CoT and LLM ensemble prompting techniques.
- We show that small-parameter models perform comparably well, and the output is considered to be aligned with human evaluations, which minimizes the need for extensive computational resources as well as lowers costs in educational settings.

The remainder of this paper is structured as follows: Section 2 reviews existing automated grading methods, highlighting their advancements and limitations. Section 3 details the framework we propose. Section 4 presents the experimental results, comparing the scores generated by our framework with those from human grading and evaluating the findings. Section 5 is the discussion, where we explore the framework's limitations and potential future research directions. Section 6 concludes the paper.

## II. BACKGROUND

Grading assignments and quizzes, especially when assessing code, can be complex and time-consuming. Code evaluation involves multiple aspects, such as code efficiency, readability, and functionality. Therefore, how educators can more effectively grade code has become noteworthy.

A survey of Automatic Assessment Tools (AATs) indicates that research related to AATs has steadily increased from 2017 to 2022 [12]. However, most of these technologies provide feedback limited to unit test results, output comparisons, or differences from reference solutions. With the advances of ML/AI, algorithmic models were introduced into automated grading systems, and representation learning [13] further simplified the grading process and improved efficiency. Before the introduction of LLMs into automated grading tasks, ML-based automated grading primarily relied on executing test cases, using techniques such as NLP, RNN, CNN, and LSTM to enhance grading accuracy [2]. With recent developments in sequence modeling and LLMs, grading capabilities expanded to include code readability and efficiency. However, few tools assess code maintainability, readability, or documentation, with most relying on static analysis (such as code quality metrics) to evaluate correctness.

In the current programming landscape, "correct" code now refers to "high-quality" code that works properly while being readable and well-structured [14]. Functionality alone isn't enough—good code must be clear, maintainable, and organized. Therefore, grading should assess correctness, readability, and functionality, as these factors together define the quality of the code.

The use of LLMs in automated grading has sparked academic interest in recent years. With their exceptional natural language processing abilities, LLMs have opened up new avenues for the field. Numerous studies have explored their applications in code grading [14], but previous research has generated feedback without validating its quality. Traditional automatic metrics (such as BLEU [15] and ROUGE [16]) show weak correlations with human judgments, leading recent studies to recommend LLM-based evaluators for assessing text generated by LLMs [11] [17]. These studies demonstrate the potential of LLMs to serve as both graders and evaluators in automated grading tasks.

## III. CODEV: LLM-BASED GRADING WORKFLOW FOR CODE EVALUATION

In introductory programming courses, educators must evaluate many programming problems and student submissions based on established grading policies and criteria. This process frequently demands substantial time and effort from educators, highlighting the importance of developing an effective automated code grading framework. To address this issue, we introduce the CodEv framework, which integrates multiple techniques to optimize grading outcomes and enhance the overall efficiency and accuracy of the assessment process.

As shown in Figure 1, the process is structured into three key stages: (1) Template design: generating templates for each problem. (Section 3.A), (2) LLMs ensemble: repeatedly generating multiple responses to perform LLM ensemble. (Section 3.B), (3) Robustness testing and result generation: using statistical methods to evaluate the robustness of the model. (Section 3.C).

### A. Template Design

Our framework develops prompt templates based on the CoT principle to guide LLMs on automated code grading effectively. The template demonstrated in Figure 2 depicts two types of prompt templates, zero-shot and zero-shot-CoT, both can incorporate custom criteria, including the correctness of output, code readability, and code functionality, for more accurate grading. For the zero-shot learning, we simply request the LLMs to generate a score and constructive feedback for each code file. For the zero-shot-CoT learning, we further request the LLMs break down the evaluation of each custom criterion into steps. By requesting the LLMs to explain the reason for the scores in each criterion step by step, we enhance the LLMs' reasoning ability to ensure more accurate grading and comprehensive feedback from different aspects defined in custom criteria.

### B. LLM Ensemble

CodEv uses ensemble techniques to improve the stability of outputs. Recent research indicates that small-parameter models perform as well as larger ones by using ensemble techniques

```
As a professional and knowledgeable expert in programming and education, you excel at reading code and can provide fair and accurate evaluations based on the given information.

Below are the problem statements and criteria; please use them to assess the code step by step.

problem statement:
[problem_statement]
 criteria:
[criteria]

   Your task is to evaluate the code based on the provided criteria. For each criterion (correctness of output, code readability, functionality), provide a step-by-step breakdown of the evaluation in smaller steps. Each criterion must have a score (e.g., 'Correctness of Output (60/80)') and an explanation of how that score was determined.

   Please respond using the following template for `reasoning_steps`, ensuring the response is a single string in narrative form:
   - `reasoning_steps`: A comprehensive text explanation in string format, following the structure below:
    "Step-by-step breakdown of the evaluation process:
    Correctness of Output (?/80): [Reason for the score]
    Code Readability (?/10): [Reason for the score]
    Functionality (?/10): [Reason for the score]"

   - `comment`: Detailed feedback on the code's performance, considering the correctness of output, code readability, and functionality, based on the `reasoning_steps`.

   - `score`: The total score (0-100) based on the evaluations.

Only generate JSON objects in your output without using a ```json block.
```

Fig. 2. This figure illustrates the template structure used in this study. Users are required to provide the problem statement and evaluation criteria to construct th assessment template. Highlighted sentences represent additional prompts derived from the CoT approach. The inclusion of the reasoning_steps structure reflects enhancements inspired by CoT, ensuring a step-by-step breakdown tailored to the criteria provided by the user. Examples include Correctness of Output (0-80), Code Readability (0-10), and Functionality (0-10)..

[18], which improve LLM performance in processing complex tasks. Based on these insights, our framework leverages LLMs to process queries based on the problem statement, evaluation criteria, and student answers, which include code and results from compiled test data, by querying each model multiple times. These queries generate results, including scores, comments, and reasoning. Then we apply a sampling and voting technique from Li et al.'s work to determine the final score of each student's code. The final score is established by identifying the most frequent score across multiple queries. This approach enhances model performance, and reduces random fluctuations in individual query results, leading to more reliable evaluations.

### C. Robustness Test and Result Generation

In the final stage of the framework, the reliability and stability of the output scores were evaluated. After conducting agreement tests, the comments accompanying these consensus answers were summarized. The consistency within and between models was also examined to assess the stability of model performance in the code evaluation task.

Inter-model and intra-model agreement tests were performed on LLMs, using the Intraclass Correlation Coefficient (ICC) as the main metric to assess scoring agreement. ICC was selected for its effectiveness in handling multiple raters and its suitability for evaluating continuous variables like scores. In contrast, other agreement metrics such as Krippendorff's Alpha and Cohen's Kappa, are more appropriate for categorical variables and designed for categorical data and are unsuitable for grading tasks. ICC quantifies the consistency between different models and within the same model across multiple ratings, providing a comprehensive understanding of stability and reliability of the ratings [19].

In this study, following Koo and Li (2016), each query was treated as a rater when a single model evaluated different tasks across multiple sessions. ICC (2, k) was used as the agreement testing metric, based on a two-way random model The model assumes that the raters (in this case, the multiple queries from a single model) are randomly selected from a larger population. As a result, the scoring reflects an element of randomness, making it generalizable to similar contexts. ICC(2, k) is therefore suitable for measuring the scoring stability of the same model under random conditions. Even if the scoring results differ from one query to another, ICC(2, k) effectively captures the agreement of the model across multiple random evaluations.

For inter-model evaluations, different models were treated as raters, and ICC(3, k) was used to assess agreement. The choice of ICC(3, k) is based on a two-way fixed model, which is appropriate for results that pertain only to a specific combination of raters—in this case, the "expert group" formed by the different models. Unlike ICC(2, k), ICC(3, k) is not generalizable to other raters. It measures consistency among specific models when evaluating the same task, meaning the results are relevant only to this particular set of models.

This dual-testing approach thoroughly analyzes the models' consistency in the proposed framework and evaluates the framework's applicability and reliability across different models.

In addition to testing the agreement of scoring results in both intra-model and inter-model contexts, we also evaluated the quality of the feedback generated by CodEv. The final feedback is derived from the evaluated results of the LLM ensemble (detailed in Section 3.B). The relevant comments, denoted as $C = \{Comment_1, Comment_2, \ldots, Comment_Q\}$, are summarized by the LLM to produce the final score.

After obtaining the final score, we used GPT-4o [20] to evaluate the quality of the generated final comments with a G-Eval-based scoring system. This method normalizes the scores by using the probabilities of output tokens from LLMs and taking their weighted summation as the final results. Formally, given a set of predefined scores (from 1 to 5) in the prompt $S = \{s_1, s_2, \ldots, s_n\}$, the probability of each score $p(s_i)$ is calculated by the LLM. The final score is determined by taking the weighted sum of $p(s_i)$ along with their corresponding scores $s_i$.

This method produces more fine-grained, continuous scores that better reflect the quality and diversity of the generated texts. Educators can use this metric to evaluate feedback from the framework. For lower scores, manual checks can be done to ensure the comments' quality and prevent the model from giving misleading or incorrect feedback to users.

TABLE I. A BRIEF INTRODUCTION TO THE SAMPLE PROBLEM USED

| Task | Description |
| --- | --- |
| Cyclic Quadrilateral Check | Determine if four given angles form a cyclic quadrilateral. |
| Sum of Factors Calculation | Compute the sum of all factors (excluding the integer itself) of a positive integer entered by the user. |
| Greatest Common Divisor (GCD) Calculation | Write a recursive C function to find the GCD of two integers using the Euclidean algorithm. |

## IV. EXPERIMENT

### A. Dataset and Benchmark

The data for this experiment were obtained from an introductory programming course attended by participants with an information technology background. Several programming test problems from the course were selected as sample questions to assess students' programming skills and problem-solving abilities. The specific requirements and evaluation criteria are outlined in Table 1. To provide a representative assessment, we collected code samples from 30 randomly selected students for each problem from the course.

In this study, we use teaching assistants' manual grading as a benchmark, reflecting human graders' standards. This allows us to assess whether the model can produce results comparable to human graders and to evaluate the framework's effectiveness in reducing grading time while maintaining accuracy and consistency.

### B. Ensemble

We consider Llama 3.1 (8B) [21] and Gemma2 (9B) [22] as the small-parameter models for experimentation and employed Llama 3.1 (70B) and GPT-4o as the large-parameter models to validate the ensemble results. The experiment compared the stability of results obtained under different ensemble sizes for each model using various methods. Specifically, we compared the sampling-and-voting method proposed in previous studies [18], the average method, and the median method. Our goal is to enable the LLM to produce scores that closely resemble those of human graders, thereby reducing the time educators spend on grading tasks. To evaluate the effectiveness of these methods, we used the Mean Absolute Error (MAE) between the predicted scores and the scores from human graders as the performance metric, which is defined in Equation 1:

$$\text{MAE} = \frac{1}{m}\sum_{i=1}^{m}\left|\text{Mode}\{\text{Score}_{i,q}\}_{q=1}^{Q} - \text{Score}_{i,\text{human}}\right| \quad (1)$$

where m represents the number of students, and $Q$ is the number of ensemble iterations for each student. $Score_{i,q}$ refers to the score of the $i$-th student during the $q$-th iteration of evaluation. $Score_{i,human}$ represents the human-graded score for the $i$-th student.

Figure 3 compares the performance of the sampling and voting method in random and worst-case scenarios across different models as the ensemble size increases. The worst-case is based on conducting multiple queries on the LLMs, sorting the results by loss values, and extracting the lowest-performing results for each ensemble size. The left side shows that most models' mean MAE remains stable or slightly decreases, indicating that the ensemble method reduces random variability. The right side shows the worst-case MAE, which decreases for larger models such as Llama3.1:70b and GPT-4o as the ensemble size increases. This indicates that larger ensembles

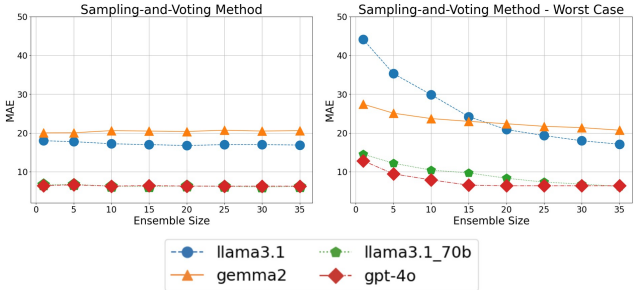

Fig. 3. The MAE performance of the ensemble method using sampling and voting with zero-shot prompts in both random and worst-case scenarios. Increasing ensemble size effectively enhances the stability of model predictions.

TABLE II. MAE OF DIFFERENT MODELS UNDER DIFFERENT ENSEMBLE METHODS WITH ENSEMBLE SIZE OF 10

| Model | Average | Sampling and Voting | Median |
|---|---|---|---|
| Llama3.1 8B | 19.02 | **17.22** | 17.42 |
| Gemma2 9B | **20.39** | 20.61 | 20.62 |
| Llama3.1 70B | 6.20 | 6.30 | **5.81** |
| GPT-4o | 6.69 | **6.30** | 6.44 |

help mitigate extreme errors. Similar trends are observed for the Median and Mean ensemble methods, confirming that ensemble techniques enhance stability and reduce high-error cases for LLMs with random outputs

Table 2 shows the MAEs of different LLMs after performing 10 ensemble iterations. The reason for choosing 10 ensemble iterations is that, under random conditions, most models have shown stable results once the ensemble size reaches 10 (as shown in the left side of Figure 3). The results indicate minimal performance differences among all methods; most models achieved a lower MAE when utilizing the Sampling and voting method. Therefore, we chose this method for this experiment's ensemble selection method.

### C. Effectiveness Validation

To validate whether our method effectively reduces the gap between small and large parameter models, we compared the MAE of scores generated under our proposed framework across all models for the same problems and student submissions against the scores obtained using the zero-shot method (see Figure 4).

The results indicate that after applying the proposed CodEv, the MAE of small-parameter models (Llama 3.1 8B, Gemma2 9B) significantly decreases. No notable differences are observed in the large-parameter models (Llama 3.1 70B, GPT-4o) (see Table 3). This finding suggests that the framework can effectively reduce the performance gap between small-parameter and large-parameter models in automated grading tasks, showcasing the potential of using small-parameter models for such tasks through prompt engineering techniques.

Furthermore, within this framework, the score difference between the large-parameter models and human evaluators is approximately six points. This discrepancy may arise from human bias, which aligns with previous research findings indicating that human evaluators tend to score higher than the scores assigned by LLMs in most cases [3]. Although forcing these models to engage in reasoning could enhance performance, they are prone to additive errors or misinterpretations, making them unsuitable for automated grading applications.

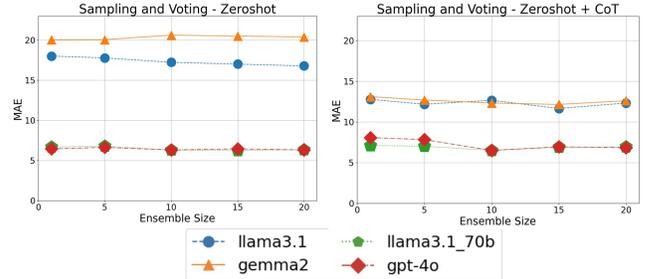

Fig. 4. Performance comparison of small-parameter models (Llama 3.1 8B, Gemma2 9B) and large-parameter models (Llama 3.1 70B, GPT-4o).

TABLE III. MAE COMPARISON OF CODEV AND ZERO-SHOT AT AN ENSEMBLE SIZE OF 10

| Model | Zero-Shot | CodEv | diff |
|---|---|---|---|
| Llama3.1 8B | 17.22 | 12.66 | -4.56 |
| Gemma2 9B | 20.61 | 12.32 | -8.29 |
| Llama3.1 70B | 6.30 | 6.50 | +0.20 |
| GPT-4o | 6.30 | 6.48 | +0.18 |

### D. Agreement Test

This study conducted intra-model and inter-model agreement tests on GPT-4o, Llama 3.1 70B, Llama 3.1 8B, and Gemma2 9B to evaluate the consistency of responses from individual LLMs over multiple iterations. The goal was to assess large language models' stability in grading tasks.

**Intra-model Agreement Test**

To validate the intra-model agreement test, we conducted 20 repeated queries for each model, using the same prompt each time. To measure the agreement of responses from a single model across multiple sessions, we chose the ICC (2, k) representing average random raters as the evaluation metric, with detailed results shown in Table 4. The results indicated that all models achieved an ICC (2, k) greater than 0.97, with p-values less than 0.001 and 95% confidence intervals between [0.96, 1]. These findings demonstrate that, regardless of the model used, the consistency of responses across multiple queries within the framework proposed in this study is very high, indicating that the framework effectively ensures the stability and consistency of the models. This further confirms the effectiveness of our method in enhancing the stability and reliability of model performance, showing that the models can produce stable and reliable outputs in response to repeated queries.

TABLE IV. RESULT OF INTRA-MODEL AGREEMENT TESTS

| Model | ICC(2,k) | P-value | CI95% |
|---|---|---|---|
| Llama3.1 8B | 0.971 | <0.001 | [0.96, 0.98] |
| Gemma2 9B | 0.998 | <0.001 | [1.00, 1.00] |
| Llama3.1 70B | 0.981 | <0.001 | [0.98, 0.99] |
| GPT-4o | 0.989 | <0.001 | [0.99, 0.99] |

TABLE V. RESULT OF INTER-MODEL AGREEMENT TESTS

|  | Estimate | P-value | CI95% |
|---|---|---|---|
| ICC (3,k) | 0.840 | <0.001 | [0.78, 0.88] |

**Inter-model Agreement Test**

In the inter-model agreement test, based on the results from Section 4.C, we employed the sampling and voting method, using an ensemble size of 10 to derive the final scores from the model outputs. Different models served as raters to evaluate the scores generated by each model across various tasks (multiple assignments). We selected ICC(3, k) as the indicator for assessing the overall agreement of the scores because this metric is particularly suited for evaluating the agreement among fixed raters (the four models used in this section) across multiple assignments. The results showed that all models' ICC(3, k) values exceeded 0.84, with p-values less than 0.001 and 95% confidence intervals between [0.78, 0.88]. These results indicate a high level of consistency in the scores assigned to automated grading tasks among the different models.

*E. Verifying the Reliability of Comments*

Based on the results outlined in Section 4.C, this study collected the reasoning steps and feedback comments generated from 10 ensemble runs. The LLM was then employed to summarize these feedback responses (Figure 5 illustrates a student code sample from Task 1 and its corresponding comment summary). The study employed G-Eval, an LLM-based evaluation metric, and used GPT-4o as the evaluating model to assess the summarized feedback with the task descriptions. A random sample of 20 comments from all models was selected for evaluation, with the results indicating that the score range for all models fell between [0.58, 0.74] (on a scale from 0 to 1). Among these, GPT-4o achieved the highest score, indicating the superior quality of its generated comments. The majority of the feedback scored above 0.5, suggesting that the comments generated by the proposed framework offer valuable insights. However, there remains significant room for improvement in the quality of the generated comments.

**Input code:**

```c
#include <stdio.h>

int main() {
    int angle1, angle2, angle3, angle4;
    int isCyclic = 1;  // Initialize as true (assuming positive angles initially)

    printf("Enter 4 angles of the quadrilateral (positive integers): ");

    // Input loop with validation
    if (scanf("%d %d %d %d", &angle1, &angle2, &angle3, &angle4) != 4) {
        printf("Invalid input. Please enter four integer values.\n");
        return 1; // Indicate error
    }

    // Check for all positive angles
    if (angle1 <= 0 || angle2 <= 0 || angle3 <= 0 || angle4 <= 0) {
        printf("The angles should be positive numbers.\n");
        return 1; // Indicate error
    }

    // Check for cyclic condition (sum of opposite angles)
    isCyclic = (angle1 + angle3 == 180) && (angle2 + angle4 == 180);

    if (isCyclic) {
        printf("YES\n");
    } else {
        printf("NO\n");
    }
}
```

**Output Comments:**

**Correctness of Output:**
The code accurately checks whether the input angles form a cyclic quadrilateral by verifying that the sum of opposite angles equals 180 degrees. It also handles invalid inputs and non-positive angle values correctly, providing informative error messages in such cases. However, it assumes that the sum of opposite angles being 180 degrees is the only requirement for a quadrilateral to be cyclic, which may not always be true.

**Code Readability:**
The code is well-structured and readable. It uses clear variable names and comments that explain each section's purpose. However, there could be more descriptive variable names for better clarity, and some comments explaining the logic behind certain parts of the code would improve its overall clarity.

**Functionality:**
The code fully meets the requirements specified in the problem statement. It checks for cyclic quadrilaterals, handles invalid inputs, and provides informative error messages when necessary. However, it lacks a prompt for re-input when an error occurs, and it does not account for non-integer inputs that could be converted to integers but do not fit the criteria for being positive integers.

**Overview Comments:**
The code demonstrates a good understanding of C programming principles and effectively solves the problem at hand. However, there are some minor improvements that can be made in terms of error handling and input validation. The code is well-structured and readable, with clear variable names and comments that explain each section's purpose. To further improve it, consider adding more comprehensive error checking for cases where non-integer inputs may occur or improving the prompt for re-input when an error occurs. Additionally, reviewing the assumption about the sum of opposite angles being 180 degrees as a requirement for a quadrilateral to be cyclic might be beneficial. Overall, the code provides an accurate solution to the problem with clear readability and good structure.

Fig. 5. The sample code from Task 1 and the comments generated by this framework were applied to the Llama 3.1 70B model.

Through the manual review of the feedback generated by different models, we observed that the models occasionally provide incorrect or overly strict evaluations, particularly in edge testing scenarios. The models tend to demand comprehensive edge tests that exceed the requirements of the problem. For instance, a problem may only ask to check if the user input is a number greater than one, but the model may critique the code for not verifying if the number is an integer. While this can indeed improve the functionality of the student's code, it surpasses the scope of the problem requirements.

Nevertheless, this study also found that models can perform more comprehensive code evaluations than manual reviews, especially regarding functionality. Since test data is typically fixed, human reviewers often assess code based solely on whether it runs correctly on the given test data. In contrast, models take into account additional execution scenarios. For example, while a code may run correctly with the provided test data, the model can identify potential errors when given inputs outside the test set (such as issues from handling odd or even numbers or when inputs exceed a certain value).

Overall, the feedback generated by the framework is valuable and offers students more suggestions for improvement. However, due to the models' occasional hallucinations in evaluation, there is still room for improvement in the quality of the feedback.

## V. Discussion

### A. Limitations

Despite our experiments showing that the CodEv framework reduces the capability gap between smaller and large parameter models, it still faces challenges in practical uses. First, the feedback quality and grading accuracy are limited by the size of the model parameters. Smaller models may occasionally produce incorrect feedback due to hallucination or being too harsh in edge cases. Secondly, although the framework focuses on code performance, experiments show that code-specific models may underperform compared to general-purpose models with similar sizes in evaluating code quality. Lastly, the results may be biased by the subjective factors in the human-graded ground truth of scores and the issue of self-judgment of GPT-4 in feedback assessment. Criteria like readability and functionality could be subjective for human evaluators, leading to inconsistencies in graded scores between the models and humans. Using GPT-4 to assess feedback generated by itself may reduce the fairness of assessment, introducing human evaluators to qualitative assessment could mitigate the problem.

### B. Future Works

This study highlights several promising directions for improving code evaluation tasks. First, while the framework helps narrow the gap between small and large parameter models, smaller models still lag behind in performance. A possible solution is to use a step-by-step distillation approach [23], where a large-parameter model trains smaller models using the reasoning steps and scores generated by the framework. This could make automated grading more affordable for educators and enable students to self-assess using smaller models on web or mobile platforms in the future.

Another important area for future research is the development of objective standards for evaluating code readability and functionality. Currently, both LLMs and human evaluators introduce subjective bias. Creating standardized evaluation criteria would improve accuracy and fairness in assessments.

Finally, there is a need to train models that excel in both reading comprehension and code understanding. Small general models often lack deep code knowledge, while specialized models struggle with reading comprehension. Future efforts should focus on developing models that balance both skills within a limited parameter size, improving the overall effectiveness of code evaluation.

## VI. Conclusion

This study introduces CodEv, an interpretable and stable framework for automated code grading that provides understandable scores and constructive feedback. It integrates various techniques to enhance LLM performance, narrowing the gap between large and small parameter models. Even small-parameter LLMs can produce scores close to human evaluators while maintaining stable results. Experimental results show that general-purpose LLMs with better reading comprehension outperform specialized LLMs in coding tasks, highlighting the importance of text understanding over specific programming knowledge. The size of the model parameter also affects grading, with large-parameter models demonstrating a better understanding of the problems and criteria, producing feedback closer to human evaluators.

Based on the experimental results, we employed the ICC for agreement testing, confirming the reliability of the generated feedback. The results indicate that LLMs, regardless of parameter size, achieve strong consistency in grading. While challenges in generating high-quality feedback remain, the framework offers educators a valuable tool for thorough code evaluation and feedback, supporting student learning. Additionally, small parameter models show promise as cost-effective options for automated code grading.